# Characterizing the Usefulness of Code Review Comments in Scientific Software for Software Quality and Scientific Rigor

Sharif Ahmed
University of Central Arkansas
Conway, AR, USA
sahmed@uca.edu

Nasir U. Eisty
University of Tennessee
Knoxville, TN, USA
neisty@utk.edu

## ABSTRACT

***Context***: Innovation thrives on scientific software, with useful code review feedback enhancing its correctness and impact. However, unlike general-purpose commercial and open-source software, the usefulness of code review feedback (*CR comment*) in scientific software remains largely unstudied. ***Objective***: This paper aims to characterize the usefulness of *CR comment* in scientific open-source software (Sci-OSS), leveraging existing research on useful *CR comment*. ***Method***: To achieve this objective, we mine successful Sci-OSS from GitHub, analyze their *CR comments* with usefulness-related features, and compare the findings from prior research on general-purpose commercial and open-source *CR comments*. ***Results***: The investigation on the usefulness of *CR comments* in Sci-OSS confirms many characteristics that prior research identified in general-purpose software. For example, subjective or negative *CR comments* remain not useful for the Sci-OSS. We also find *CR comments* which receive negative emoji reactions have a very small correlation with not useful comments, whereas the positive emojis show mixed correlations. Importantly, 6-33% *CR comments* in Sci-OSS are not useful in our mined repositories. ***Conclusions***: Our investigation into Sci-OSS extends findings from *CR comments*' usefulness research on general-purpose software, benefiting developers, scientists, and researchers in the Sci-OSS community.

## 1 INTRODUCTION

Modern Code Review (MCR), online asynchronous tool-based source code inspection technique, is crucial despite its significant time costs [13]. According to Bosu et al. [7], developers spend at least a tenth of their time engaging in peer code review activities in both industry and open-source development. Given its essential role, the code review process is beneficial in enhancing overall team efficiency and project quality. The core of the MCR process lies in the *CR comments* [13].[1] *CR comments* are different from source-code comments and commit messages that are written by code-authors instead of code-reviewers. The effectiveness of the code review process is heavily dependent on the usefulness of these *CR comments*. Usefulness of *CR comments* also emerged in Kononenko et al. [18]'s empirical study on the Mozilla code review process. Despite their importance, empirical findings from Bosu et al. [8] reveal that more than 1 out of 3 *CR comments* at Microsoft are not useful.

> `Given enough eyeballs, all bugs are shallow`
> —Linus Torvalds [28]

This well-known quote captures the core idea behind open source software (OSS): when many people can view, test, and improve code, problems are more likely to be found and fixed quickly. OSS relies on collaboration, transparency, and community involvement to ensure software quality. While many open source tools are built for general use, such as web browsers, operating systems, or development frameworks, some are created specifically to support scientific research. These are known as scientific open-source software (Sci-OSS). They help researchers analyze data, run experiments, and test new ideas. Although both types of software are open and collaborative, they serve different goals, follow different development practices, and are used by different communities. Moreover, different people, groups, and organizations write and maintain the Sci-OSS. These people may have different expertise, backgrounds, and practices. However, researchers have investigated the usefulness of code review feedback or code review comments (*CR comments*), identified factors to classify usefulness of *CR comments* [8, 31]. We also featurized, predicted, compared and contrasted the usefulness of *CR comments* between commercial and open-source general software in our previous study [3].

Similar to general software, code review is crucial in Sci-OSS because it helps ensure the accuracy, reliability, and trustworthiness of computational results, which are fundamental to scientific progress [16]. Given that Sci-OSS often involves complex algorithms and data analysis, peer review of code allows developers to identify and correct errors, improve readability, and enhance maintainability. This collaborative process not only reduces the likelihood of bugs

[1] "*CR comments*" and "Code Review Comments" are interchangeably used in this paper

and inaccuracies in scientific outputs but also promotes knowledge sharing among developers. As a result, implementing systematic peer code review practices contributes to producing high-quality Sci-OSS that can be trusted by the scientific community and supports reproducible and robust scientific outcomes.

To the best of our knowledge, none of the prior research has framed the usefulness of *CR comments* in Sci-OSS. This observation motivates us to look into Sci-OSS, leveraging state-of-the-art techniques for identifying and analyzing *CR comments*. **This paper aims to investigate the usefulness of *CR comments* in Sci-OSS.** It complements our previous work [3] and other existing research conducted in commercial and open-source *CR comments* [24, 27]. Analyzing the state-of-the-art estimated usefulness of *CR comments* in Sci-OSS via feature analysis and eXplainable AI (XAI) can inform us what makes a *CR comment* useful in Sci-OSS.

The outcome of the proposed work is based on addressing the following research questions:

- To the best of our knowledge, there is no dataset or experiment on usefulness of *CR comments* in Sci-OSS. So, we ask first research question, **RQ1:** How useful are the Sci-OSS *CR comments* when evaluated with state-of-the-art usefulness prediction models?
- We have found studies that characterize and describe usefulness of *CR comments* in general software [3, 8, 31]. To describe the *CR comments* in Sci-OSS leveraging usefulness research we ask our second research question, **RQ2:** How do the characteristics of *CR comments* in Sci-OSS differ from those of useful *CR comments* in commercial and open-source?
- Our recent work [3] observed bipolarity of usefulness characteristics in *CR comments* from different environment. Therefore we pose our third research question, **RQ3:** Do *CR comments* in Sci-OSS projects vary in characteristics across different scientific domains?
- Our other work [2] found emoji-features help predicting usefulness of *CR comments* better, thus we explore our last research question, **RQ4:** How are emojis perceived in *CR comments* within Sci-OSS?

The rest of the paper is organized as follows. Section 2 provides background and a brief overview of the literature. Section 3 details the methodology for answering our research questions, **RQ 1-4**. Section 4 reports and discusses our experiment results. Section 5 reviews the threats to the validity of our work, and Section 6 concludes.

## 2 BACKGROUND AND RELATED WORK

This section provides background and a concise review of the relevant literature.

### 2.1 Usefulness Definition

In 2014, Pangsakulyanont et al. [25] proposed a definition for *CR comments*, classifying them as either *useful* or *useless* based on how closely the reviewer's comment matched the author's commit message for a specific code change. If it was not possible to clearly determine the usefulness of a comment, it was labeled as *undetermined*. The following year, Bosu et al. [8] studied *CR comments* at Microsoft, where seven developers labeled them as *useful*, *somewhat useful*, or *not useful*. For their usefulness prediction model, *somewhat useful* comments were grouped with *useful* ones. A *CR comment* was considered *useful* if it led to a code change within 1–10 lines of the comment. They found that changes one line away minimized both *false positives* and *false negatives*. This definition was later adopted [24, 27] or adapted [31] in three datasets.

### 2.2 Classifying Usefulness

Initially, Pangsakulyanont et al. [25] employed the Vector Space Model with cosine similarity measure between the commit message and the *CR comment* on a code-change to classify usefulness. Subsequently, Bosu et al. [8] identified the factors of useful *CR comments* through an empirical study at Microsoft. To generate features for their usefulness classifier, they utilized textual properties of *CR comments* and attributes of review activities. Kononenko et al. [19] introduced factors, such as code author and reviewer experience, that impact the quality of code reviews. Rahman et al. [27] collected *CR comments* from commercial projects and developed a model to predict their usefulness. Their approach involved using textual properties of the *CR comments* and features related to the developers' experience. However, unlike Bosu et al. [8], they did not incorporate any features related to the code review activities. In 2018, Efstathiou and Spinellis [14] proposed measuring usefulness using linguistic semantics to address this issue. Concurrently, Meyers et al. [24] used several linguistic features to classify *CR comments*. Hasan et al. [17] developed an in-house web-based application to reward reviewers for their useful feedback. They did not consider any linguistic features from Meyers et al. [24]. However, they extended the existing features by incorporating features related to code review activities. Furthermore, they presented several features for predicting the usefulness of *CR comments* from the literature and their work, but these features are limited to their model.

For the feature-based usefulness prediction task, Rahman et al. [27] utilized the *Mann-Whitney Wilcoxon test* to identify differences in features between useful and non-useful *CR comments*, and *Cohen's D* as an effect-size measure. In contrast, Hasan et al. [17] used the *Pearson Correlation* measure to interpret their input features. Turzo and Bosu [31] employed the Chi-Square test to check the relationship between respondents' demographics and responses. Recently, we [3] analyzed the features by combining feature analyses from prior works [17, 27]. In that work, our comprehensive feature analyses on cross-datasets uncovered a bipolarity between datasets of *CR comments* from two different environments.

### 2.3 Emojis and Usefulness

Emojis and emoticons serve as non-verbal cues and convey emotive or instructive messages to developers. In one of our previous studies [2], we explored the utility of *CR comments* by examining the sentiments and semantics of emojis within these comments using existing datasets $\mathcal{D}_{rh}$,$\mathcal{D}_{cc}$, and $\mathcal{D}_{od}$. In that paper, emoji-aware models outperform the emoji-unaware models. Therefore, overseen emojis in *CR comments* emerged as an important aspect of *CR comments*' usefulness prediction task.

## 2.4 General *CR comments*

In another of our previous works [1], we identified three available datasets of useful *CR comments* from general software. We briefly describe them below.

- $\mathcal{D}_{rh}$, **2017:** This dataset [27] comprises 1,481 *CR comments*, collected through the GitHub API from four commercial projects of an unspecified company. The authors of the article manually annotated each *CR comment* as either *useful* or *non-useful*, based on the usefulness heuristic of the Microsoft study [8].
- $\mathcal{D}_{cc}$, **2018:** This dataset [24] has 3,794 *CR comments* from Google Chromium Project. The authors used the RESTful API to obtain accessible *CR comments* on Rietveld (code review tool used in Google Chromium Project) from 2008 to 2016. Next, they automatically identified acted-upon *CR comments* by considering Rietveld's "Done" click feature. Though their annotation principle is similar to the $\mathcal{D}_{rh}$ [27], it was labeled as *acted-upon* (i.e., *useful*) and *not*-(known to be)-*acted upon CR comments*.
- $\mathcal{D}_{od}$, **2023:** This dataset [31] contains 2,654 *CR comments* from OpenDev's Nova OSS project. The authors used the Gerrit Miner tool for mining *CR comments* from 2011 to 2022 and manually annotated *CR comments* with *useful* and *not-useful* labels.

In this paper, we use all three general-purpose commercial and open-source *CR comment* datasets ($\mathcal{D}_{rh}$,$\mathcal{D}_{cc}$,& $\mathcal{D}_{od}$) to compare our *CR comments* derived from SciOSS, $\mathcal{D}_{sci}$.

## 2.5 SciOSS *CR comments*

The U.S. Department of Energy[2] (DoE) sponsored CASS[3] fosters collaboration among diverse Software Stewardship Organizations (SSOs), each responsible for advancing parts of the Sci-OSS ecosystem (e.g., math libraries, data tools). Ahmed et al. [4] explored the scientific open-source software (e.g., math libraries, data tools) and mined 10 projects that are open and active on GitHub. In the process of project selection, authors excluded the projects that are mirrored on GitHub from social coding or other platforms.

Next, authors mined the GitHub data using GitHub REST API. Their dataset contain other information in addition to *CR comments*, such as, pull request number, line number, author association, user type, and GitHub emoji reactions (👍👎😕❤️🚀👀🎉😄). We name these additional data related to the mined *CR comments* as *metadata* throughout this paper.

To make our work self-contained, we provide brief overview of approximated and normalized metadata [4], which we use to answer one of our research questions.

*Approximated Data.* GitHub user profiles do not have specific attributes for the developer's gender, but gender diversity has been studied and found to be an important metric for community engagement and software sustainability. Ahmed et al. [4] employed an open-source gender-guessing technique [12] that is able to predict gender nearly 98% of the time correctly [29]. The tool is able to predict a person's gender from their name using the categories *male, mostly male, female, mostly female, androgynous,* or *unknown.* After determining the guessed gender, they transformed "mostly male" and "mostly female" into male and female respectively, and unpredicted genders into the "unknown" category.

[2] http://science.energy.gov
[3] www.cass.community

*Normalized Data.* Ahmed et al. [4] observed the GitHub users' location data and found that a specific location had multiple values for example 'US', 'USA', 'U.S.', or 'United States'. Additionally, some GitHub users shared city names without their country names, for example, 'London' instead of 'London, United Kingdom.' To unify the geo-location of the developers, they used GeoPy [20], an open-source geocoding python client, to obtain a full address from the OpenStreetMap Nominatim geocoder. Lastly, for maximum coverage, they took only the country name from the obtained address as location data for every developer who publicly shared their location information at the time of our GitHub data mining.

We derive usefulness-related features from literature [3, 8, 24, 27] to characterize the usefulness of these SciOSS *CR comments* and contribute $\mathcal{D}_{sci}$(as detailed in Section 3).

## 3 METHODOLOGY

Here, we outline our proposed methodology to answer our research questions, **RQ1-4**, introduced in Section 1. Figure 1 shows the proposed approach for this research objective. As our focus is on the scientific projects, we first select the relevant projects. Next, we collect the data from selected projects. We then answer our research questions through our feature and XAI analyses. We detail each phase below.

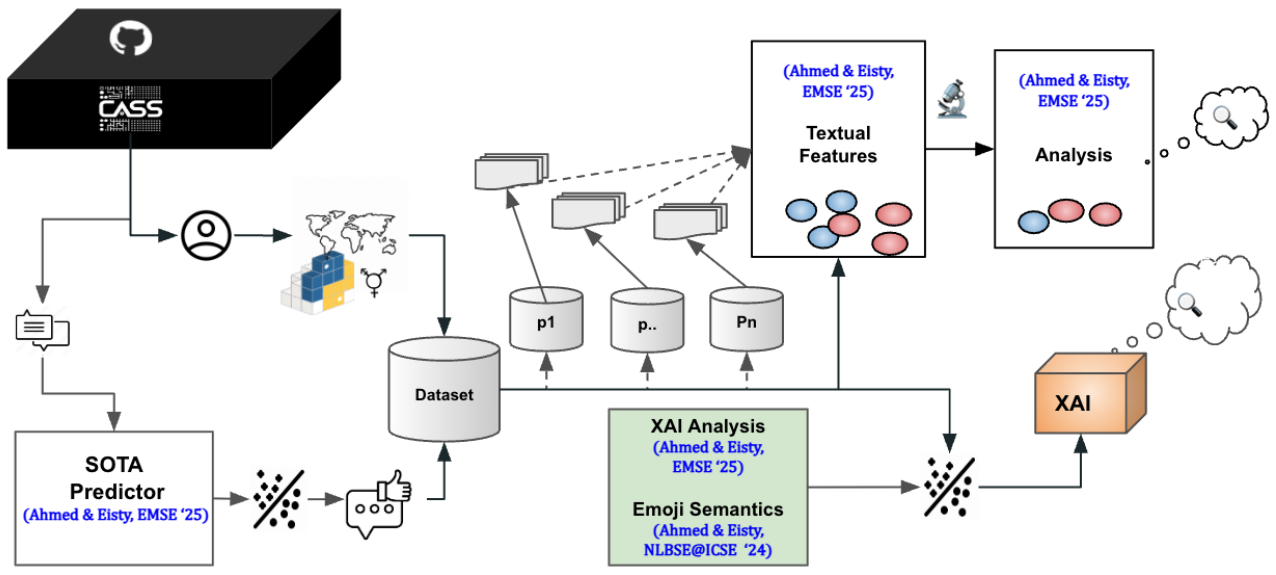

**Figure 1: Overview of Our Methodology to Answer RQs 1-4**

## 3.1 $\mathcal{D}_{sci}$ Dataset Curation

We adopt *CR comments* from SciOSS [4], compute usefulness-related hand-crafted features from these *CR comments*, and answer our research questions **RQ1-4**.

*3.1.1 Feature Data Computation.* After collecting the data, we compute the feature values using existing tools and artifacts from our prior work and related research[1–3, 8, 24, 27]. With the obtained data, we compile a dataset for our experiments and contribute it to future research. Importantly, as we are interested in studying the usefulness of scientific *CR comments* and we have state-of-the-art usefulness predictors, we plan to obtain the predictor-based annotation of our derived scientific *CR comments*. This will facilitate us to characterize the scientific *CR comments*.

#### 3.1.2 *Prediction-based Usefulness Annotation.*

This paper aims to investigate scientific *CR comments* within the context of code review feedback usefulness and characterize scientific *CR comments* leveraging features and analyses applied for understanding the *CR comments*. As we have a state-of-the-art usefulness predictor from our previous work [3], we plan to obtain a predictor-based annotation of our derived Sci-OSS *CR comments*. This paper shows OpenAI's GPT-4o as the best-performing model to predict the usefulness of code review comments. As GPT-4o is not free and our derived code reviews are many (164,708 *CR comments*) compared to their adopted datasets (<4,000 *CR comments*), we will need to pay for the fine-tuning process and then for predicting our scientific code review text. To achieve this, we choose the second-best performing model from our paper [3], which is free of cost and demands less computational time. Finally, we use the usefulness prediction model [3] to predict the usefulness of our derived Sci-OSS *CR comments*.

## 3.2 Answering Research Questions

We leverage the experiments from prior research [2, 3, 17, 27] to answer the research questions for scientific *CR comments*.

#### 3.2.1 ***RQ1****: How useful are the Sci-OSS CR comments when evaluated with state-of-the-art usefulness prediction models?*

To answer this question, we check the distributions of predicted usefulness of *CR comments* from our compiled dataset, $\mathcal{D}_{sci}$. For better insight, we plot the aggregated usefulness ratio of the individual projects within $\mathcal{D}_{sci}$ over the years.

#### 3.2.2 ***RQ2****: How do the characteristics of CR comments in Sci-OSS differ from those of useful CR comments in commercial and open-source?*

Conducting feature and XAI analyses from prior *CR comments* analyses [3, 27] on Sci-OSS projects can describe the *CR comments* in Sci-OSS. Thus, we perform the existing experiments with our newly obtained scientific *CR comments*.

*A.* For **within-dataset feature analyses**, we follow the following process.

- We adopt around fifty textual features from our previous works [2, 3], which includes prior works [1, 8, 17, 24], in our experiment.
- Next, we preprocess and calculate all the feature values of our $\mathcal{D}_{sci}$ *CR comments* using our previous implementations and details [2, 3].
- With the computed feature values for each *CR comment*, we determine whether each of the individual features can distinguish auto-annotated usefulness. To this end, we employ existing statistical measures, the *Mann-Whitney U test* [23] and *Cohen's D* [10], that were used for inspecting usefulness-related feature analysis [3, 27].
- We compare and contrast the statistical findings from our scientific *CR comments* with findings from general-purpose commercial and open-source *CR comments*.

*B.* For **cross-dataset feature analyses**, we adhere to the following steps, similar to our previous study [3]. While our within-dataset feature analysis describes the individual feature's ability to distinguish usefulness within a single dataset, the cross-dataset feature analysis evaluates whether the correlation between features and usefulness statistically differs from one dataset to another.

- Compute Pearson correlation of all the features to the predicted usefulness labels for all datasets.
- Take feature correlations on our dataset versus feature correlations on each of the existing datasets and perform the *Wilcoxon signed-rank test* [32] followed by Cohen's D [10] effect size measure.

We have noticed the bipolarity of two open-source datasets in our prior work's empirical results [3]. This answer from the Sci-OSS will inform us if the Sci-OSS has similar polarity or not.

*C.* For **cross-dataset eXplainable AI (XAI) analysis**, we adopt the recent XAI analyses on *CR comment*-usefulness datasets [3]. This approach provide explanation of a dataset when trained on an explainer AI model. To this end we take following steps.

- We consider robust SHapley Additive exPlanation framework [22] to train one of the dataset on a linear explainer model and get explanation while predicting another dataset.
- Train the selected explainer model with our curated $\mathcal{D}_{sci}$ dataset and check explanation from predicting usefulness of existing datasets ($\mathcal{D}_{rh}$, $\mathcal{D}_{cc}$, $\mathcal{D}_{od}$) and vice versa.
- The obtained SHapely values will provide what contributed in explainer model to predict a *CR comment* useful or not-useful.
- The XAI findings from scientific *CR comments*of $\mathcal{D}_{sci}$ will complement the findings of general purpose *CR comments* from commercial $\mathcal{D}_{rh}$ and open-source $\mathcal{D}_{cc}$ and $\mathcal{D}_{od}$ datasets.

#### 3.2.3 ***RQ3****: Do CR comments in Sci-OSS projects in Sci-OSS projects vary in characteristics across different scientific domains?*

To answer this RQ, we select our derived *CR comments* in $\mathcal{D}_{sci}$ that come from similar and different scientific domains, such as applied math, programming systems, machine learning. Next, we compare their usefulness and usefulness-feature values using statistical tests following our experiment setup for cross-dataset analysis in **RQ2** (Section 3.2.2). Here, we compare the projects within $\mathcal{D}_{sci}$ only.

As our dataset has additional metadata from the Sci-OSS repositories (Sec 2.5), we compare and contrast these metadata in terms of *CR comment* usefulness (Sec 3.1.2). Here, we conduct statistical-tests for feature data (described in Section 3.2.2) on these metadata. We also consider Pearson correlation ($\delta$) between these numeric metadata of *CR comments* and usefulness *CR comments*. Since some of these metadata are categorical, we consider Chi-Square ($\chi^2$) [26] instead of Mann-Whitney U and Cramér's $V$ [11] instead of Cohen's D and Pearson correlation. Answers to this will provide an additional snapshot of the selected Sci-OSS projects.

#### 3.2.4 ***RQ4****: How are emojis perceived in CR comments within Sci-OSS?.*

We examine the emoji sentiments and emoji semantics within our derived Sci-OSS *CR comments*. *CR comment*-usefulness literature informs us that emoji semantics are perceived differently in useful *CR comments*for general purpose open and commercial software [2].

- We examine which emojis are used in scientific *CR comments* comparing the literature.

- Next, we preprocess, extract, and normalize emoticons, emoji-expressions, and Unicode emojis to a single format, Unicode emojis.
- With unified emojis, we obtain emoji sentiment features and emoji semantic embeddings using Novak's general sentiment [21], Ahmed & Eisty's code review sentiments [2], and emoji2vec pre-trained language model [15].
- We then check the features' correlation to usefulness in emoji-containing *CR comments* in $\mathcal{D}_{sci}$.
- Finally, we contrast the emoji not-aware *CR comment*-semantics (fastText [6]) and emoji aware *CR comment*-semantics (fastText [6]+emoji2vec [15]) for predicting the usefulness of *CR comments*. We maintain the existing stratified 10-fold cross-validation setup for this evaluation [2]. Since our dataset has prediction-based annotation, we limit our analysis for prediction-evaluation to $\mathcal{D}_{sci}$. We additionally employ a statistical paired comparison between the $\mathcal{D}_{sci}$'s annotated usefulness and emoji-(non)-aware models' *CR comment*-usefulness predictions. For this measure, we consider Wilcoxon's Signed Rank test [32] with a threshold of p-value$< 0.05$, followed by Cohen's D [9] effect-size measure.

This research question will extend our knowledge by adding insights for emojis within the context of the Sci-OSS code review feedback.

# 4 RESULTS AND DISCUSSIONS

This section provides the experimental results corresponding to the procedures described in Section 3 and offers analysis of findings.

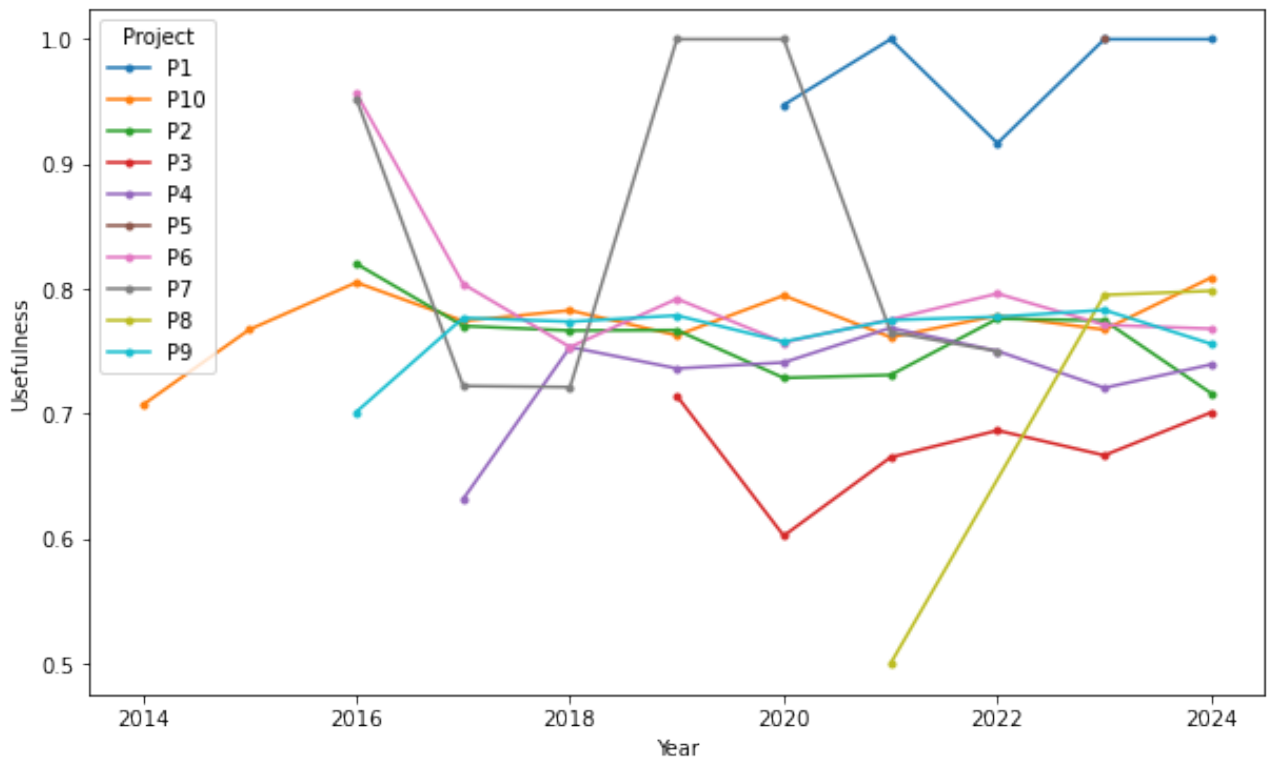

Here, y-axis shows yearly aggregated ratio (0–1.0) of useful *CR comments* across 10 SciOSS projects (2014-2024)

**Figure 2: Useful-*CR Comments* in Scientific Software, $\mathcal{D}_{sci}$**

## 4.1 RQ1: Usefulness of *CR comments* in Sci-OSS

We find 79% of our derived 164,708 *CR Comments* in $\mathcal{D}_{sci}$ are useful with our prediction based useful annotation described in Section 3.1.2. The project-wise ratio of not-useful *CR comments* in $\mathcal{D}_{sci}$ ranges from 6% to 33%. This mirrors the findings of a study conducted a decade ago at Microsoft [8], which reported 34.5% of *CR comments* were not-useful.

Figure 2 shows the yearly aggregated ratio of useful comments of the scientific projects. Most of the projects had 70-80% useful *CR comments*. The project P5 has only one datapoint in the figure with 100% useful *CR comments*. After looking into this, we find that this project has only 19 *CR comments* and all of them are from the year 2023. For P8, the useful comments ratio in year 2021 is the lowest, 50% useful from only 10 *CR comments*. However, it has 80% useful *CR comments* in the following years, 2023 with 32,205 *CR comments* and 2024 with 71,150 *CR comments*.

## 4.2 RQ2: Useful *CR comments*: General vs Sci-OSS

*4.2.1 Features.* Table 1 shows statistical properties of scientific *CR comments* from our $\mathcal{D}_{sci}$ along with all three existing usefulness labeled *CR comments*-datasets, $\mathcal{D}_{rh}$, $\mathcal{D}_{cc}$, and $\mathcal{D}_{od}$, from general software. Ahmed & Eisty discovered a bipolarity between $\mathcal{D}_{cc}$ and $\mathcal{D}_{od}$, so we were curious about $\mathcal{D}_{sci}$'s stance, and we find that $\mathcal{D}_{sci}$ is statistically closer to $\mathcal{D}_{cc}$. Overall, we see that *CR comments* from Sci-OSS has a stronger association with usefulness for code review sentiment, *cr_senti*, [5] and association with not-usefulness for *stop-word-ratio, is confirmatory?* than general open-source and commercial software.

*4.2.2 Datasets.* Table 2 shows that all the projects are significantly different from each other, with Wilcoxon's signed rank p-value<0.05, except for $\mathcal{D}_{sci}$-$\mathcal{D}_{rh}$. The Cohen's D effect-size is also shows $\mathcal{D}_{sci}$ strongly varies with $\mathcal{D}_{od}$, similar to $\mathcal{D}_{rh}$ and $\mathcal{D}_{cc}$.

*4.2.3 eXplainable AI.* Figure 3 shows the words that contributed to predicting usefulness of *CR comments* from scientific $\mathcal{D}_{sci}$ to general $\mathcal{D}_{rh}$, $\mathcal{D}_{cc}$, and $\mathcal{D}_{od}$, and vice versa. This extends the findings from the recent XAI findings [3]. In our XAI analysis, we find that our XAI analysis confirms prior works regarding the contributions of the words *line, we, you, good, should, test*, etc. We also don't notice the *docstring, API, instance, method, case* words that had contributed to XAI analysis across general software in prior work [3]. Therefore, *CR comments* that specify the source-code line, avoid subjective feedback, and talk about testing are also helpful in Sci-OSS.

Following the excerpting of prior works [3, 30], we explore $\mathcal{D}_{sci}$ comments labeled as useful and present a few excerpts below.

- ‘`check history of this line`’
- ‘`This is not the focus of this PR. but the indentation of this line is confusing.`’
- ‘`nit: trivial braces can be omitted`’
- ‘`Presumably this line will change depending on the Fortran compiler used. Users will just have to be aware of that.`’
- ‘`Yep, there should be a `xxxx.c` which needs the cuda header instead`’
- ‘` Use xxxx_error_w_msg instead.`’
- ‘`@xx @xx I'd avoid including `.hpp` in a `.h` but instead include in `.c` that needs CUDA`’
- ‘`acknowledged.`’
- ‘`unresolving, @xx should take another look here and acknowledge the solution explicitly`’

**Table 1: Statistical Properties of Features in $\mathcal{D}_{sci}$ *CR comments*, Comparing $\mathcal{D}_{rh}$, $\mathcal{D}_{cc}$, and $\mathcal{D}_{od}$ from [3]**

Here, effect-sizes denoted by $*$ are statistically significant (p-value<0.05)

| | effect-size$^{\text{p-value}}$ | | | | Pearson Correlation($\delta$) | | | |
|---|---|---|---|---|---|---|---|---|
| *Dataset*⟶ | $\mathcal{D}_{sci}$ | $\mathcal{D}_{rh}$ | $\mathcal{D}_{cc}$ | $\mathcal{D}_{od}$ | $\mathcal{D}_{sci}$ | $\mathcal{D}_{rh}$ | $\mathcal{D}_{cc}$ | $\mathcal{D}_{od}$ |
| tone | **0.03**$^*$ | 0.05 | 0.17$^*$ | 0.46$^*$ | **0.01** | -0.03 | 0.07 | -0.17 |
| distress | **0.01**$^*$ | 0.01 | 0.11$^*$ | 0.23$^*$ | **0.00** | 0.00 | 0.04 | -0.09 |
| empathy | **0.01**$^*$ | 0.01 | 0.11$^*$ | 0.23$^*$ | **0.00** | 0.00 | 0.04 | -0.09 |
| is_toxic | **0.01** | 0.06 | 0.02 | 0.19$^*$ | **0.00** | -0.03 | 0.01 | -0.07 |
| gratitude | **0** | 0.01 | 0.24$^*$ | 0.15$^*$ | **0.00** | -0.01 | -0.10 | -0.06 |
| cr_senti | **0.29**$^*$ | 0.12$^*$ | 0.19$^*$ | 0.09 | **0.12** | 0.06 | 0.08 | -0.04 |
| kw_refact_xerox | **0.15**$^*$ | 0.04 | 0.17$^*$ | 0.09 | **0.06** | 0.02 | 0.07 | -0.04 |
| num_propernouns | **0.01**$^*$ | 0.09 | 0.02 | 0.05 | **0.00** | -0.04 | 0.01 | -0.02 |
| num_interjections | **0** | 0.04 | 0.07 | 0.04 | **0.00** | 0.02 | -0.03 | -0.01 |
| polarity | **0.27**$^*$ | 0.04 | 0.25$^*$ | 0.03 | **-0.11** | -0.02 | -0.10 | -0.01 |
| kw_msoft_u | **0.21**$^*$ | 0.07 | 0.38$^*$ | 0 | **0.09** | 0.03 | 0.16 | 0.00 |
| num_exclamation | **0.01**$^*$ | 0.01 | 0.02 | 0.01 | **0.01** | -0.01 | -0.01 | 0.00 |
| is_confirmatory | **0.22**$^*$ | 0.04 | 0.07 | 0.02 | **0.09** | 0.02 | -0.03 | 0.01 |
| question_ratio | **0.19**$^*$ | 0.2$^*$ | 0.51$^*$ | 0.03 | **-0.08** | -0.10 | -0.21 | 0.01 |
| kw_satd_potdar | **0.02**$^*$ | 0.06 | 0.1$^*$ | 0.04 | **0.01** | -0.03 | 0.04 | 0.02 |
| implicature | **0.09**$^*$ | 0.01 | 0.06$^*$ | 0.06 | **-0.04** | 0.00 | -0.02 | 0.02 |
| kw_msoft_nu | **0.21**$^*$ | 0.1$^*$ | 0.26$^*$ | 0.06 | **-0.09** | -0.05 | -0.10 | 0.02 |
| code_word_ratio | **0.2**$^*$ | 0.15$^*$ | 0.23$^*$ | 0.06$^*$ | **0.08** | 0.07 | 0.10 | 0.02 |
| politeness | **0.14**$^*$ | 0.12 | 0.19$^*$ | 0.06$^*$ | **-0.06** | -0.06 | -0.08 | 0.02 |
| num_Qmark | **0.12**$^*$ | 0.18$^*$ | 0.52$^*$ | 0.1$^*$ | **-0.05** | -0.09 | -0.21 | 0.04 |
| rd_text | **0.21**$^*$ | 0.08 | 0.19$^*$ | 0.14$^*$ | **-0.08** | -0.04 | -0.08 | 0.06 |
| kw_refact_problem | **0.01** | 0.09 | 0.09 | 0.16$^*$ | **-0.01** | 0.05 | 0.03 | 0.06 |
| avg_chars | **0.2**$^*$ | 0.11$^*$ | 0.16$^*$ | 0.16$^*$ | **0.08** | 0.05 | 0.07 | 0.06 |
| formality | **0.1**$^*$ | 0.01 | 0.15$^*$ | 0.18$^*$ | **0.04** | 0.00 | 0.06 | 0.07 |
| subjectivity | **0.36**$^*$ | 0.15$^*$ | 0.38$^*$ | 0.18$^*$ | **-0.15** | -0.07 | -0.15 | 0.07 |
| has_out_snippet | **0.04**$^*$ | 0.02 | 0.09$^*$ | 0.2$^*$ | **0.02** | -0.01 | -0.04 | 0.08 |
| kw_secdev | **0.12**$^*$ | 0.01 | 0.31$^*$ | 0.24$^*$ | **-0.05** | -0.01 | -0.13 | 0.09 |
| avg_punct | **0.1**$^*$ | 0.12$^*$ | 0.04 | 0.25$^*$ | **0.04** | 0.06 | 0.02 | 0.10 |
| num_sent | **0.21**$^*$ | 0.02 | 0.46$^*$ | 0.25$^*$ | **-0.09** | 0.01 | -0.19 | 0.10 |
| stop_word_ratio | **0.37**$^*$ | 0.02 | 0.24$^*$ | 0.26$^*$ | **-0.15** | -0.01 | -0.10 | 0.10 |
| num_tentative | **0.17**$^*$ | 0.08 | 0.23$^*$ | 0.27$^*$ | **-0.07** | -0.04 | -0.10 | 0.10 |
| informativeness | **0.06**$^*$ | 0.02 | 0.07 | 0.28$^*$ | **0.02** | 0.01 | 0.03 | 0.11 |
| programming_words | **0.02** | 0.02$^*$ | 0.12 | 0.31$^*$ | **0.01** | -0.01 | -0.05 | 0.12 |
| num_adverb | **0.27**$^*$ | 0.01 | 0.39$^*$ | 0.32$^*$ | **-0.11** | -0.01 | -0.16 | 0.12 |
| num_adj | **0.13**$^*$ | 0.16$^*$ | 0.36$^*$ | 0.33$^*$ | **-0.05** | -0.08 | -0.15 | 0.12 |
| num_nouns | **0.06**$^*$ | 0.07 | 0.39$^*$ | 0.35$^*$ | **-0.02** | -0.04 | -0.16 | 0.13 |
| num_determinants | **0.26**$^*$ | 0.09 | 0.43$^*$ | 0.37$^*$ | **-0.11** | -0.04 | -0.17 | 0.14 |
| num_chars | **0.11**$^*$ | 0.03 | 0.46$^*$ | 0.39$^*$ | **-0.05** | -0.02 | -0.18 | 0.15 |
| num_words | **0.2**$^*$ | 0.07 | 0.5$^*$ | 0.39$^*$ | **-0.08** | -0.03 | -0.20 | 0.15 |
| avg_stopwords | **0.29**$^*$ | 0.1 | 0.38$^*$ | 0.4$^*$ | **-0.12** | -0.05 | -0.15 | 0.15 |
| kw_refact_solution | **0.19**$^*$ | 0.01 | 0.41$^*$ | 0.41$^*$ | **-0.08** | 0.00 | -0.17 | 0.16 |
| num_verb | **0.16**$^*$ | 0.03 | 0.38$^*$ | 0.41$^*$ | **-0.07** | 0.02 | -0.15 | 0.16 |
| avg_words | **0.11**$^*$ | 0.08 | 0.32$^*$ | 0.44$^*$ | **-0.04** | -0.04 | -0.13 | 0.17 |

**Table 2: Comparisons among Features' Correlations to the Usefulness**

Here, $*$ indicates statistically significant (p-value<0.05)

| | | Cohen's D effect-size | | | |
|---|---|---|---|---|---|
| | | $\mathcal{D}_{cc}$ | $\mathcal{D}_{od}$ | $\mathcal{D}_{rh}$ | $\mathcal{D}_{sci}$ |
| p-value | $\mathcal{D}_{cc}$ | | 1.28 | 0.64 | 0.51 |
| | $\mathcal{D}_{od}$ | * | | 1.08 | 1.28 |
| | $\mathcal{D}_{rh}$ | * | * | | 0.24 |
| | $\mathcal{D}_{sci}$ | * | * | - | |

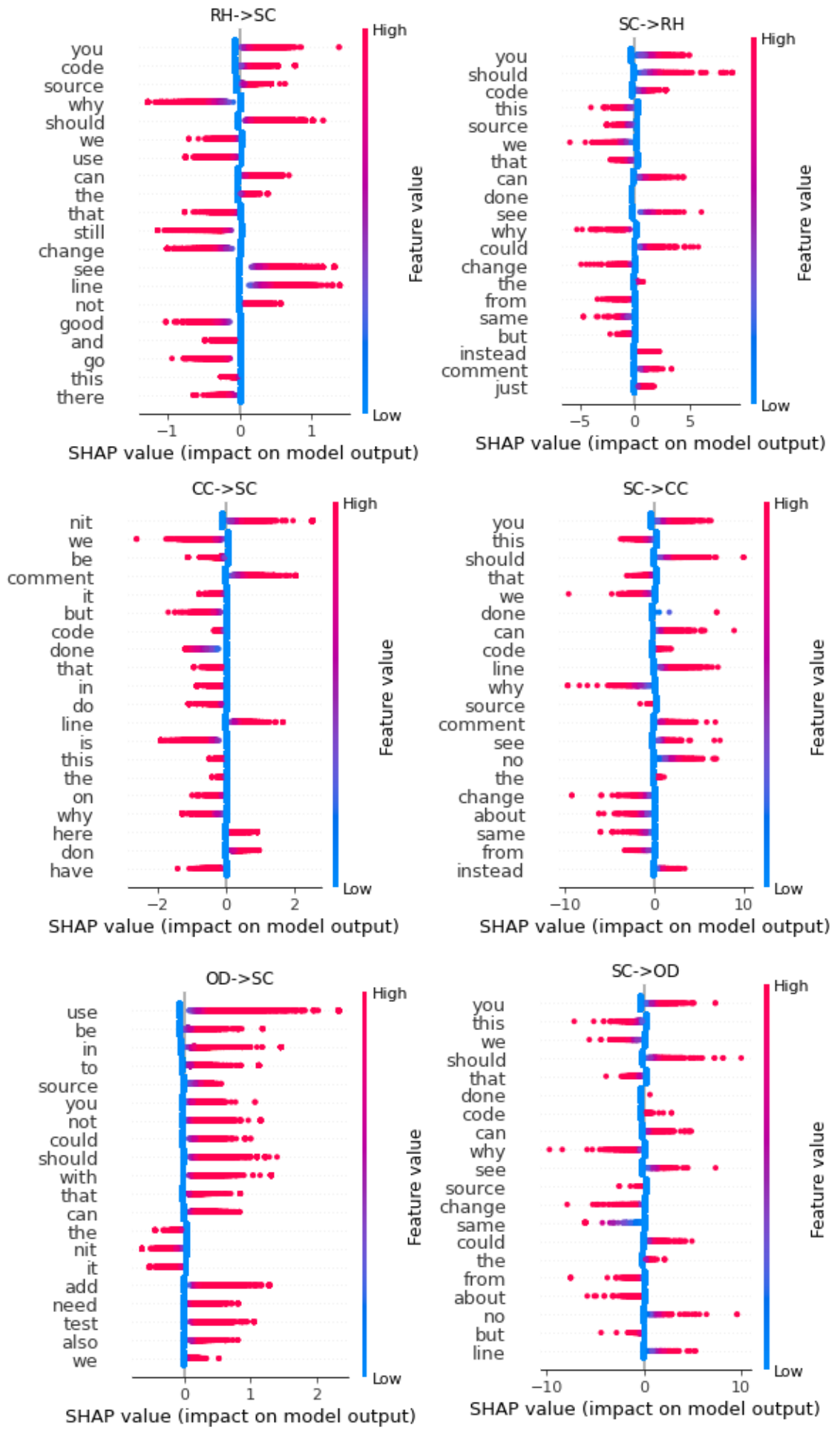

**Figure 3: Interpreting Scientific and General *CR comments* using XAI**

The excerpts above from $\mathcal{D}_{sci}$ have similar instructions or messages that are also useful in general *CR comments*, such as refactoring, suggestions, and nitpicking [3].

## 4.3 RQ3: Useful *CR comments* in Different Scientific Domains

We report analysis from both features' data, computed from the *CR comment*, and the metadata, derived from the GitHub repository attributes associated with *CR comments*, from our compiled $\mathcal{D}_{sci}$.

*Metadata.* Table 3 shows the statistical relationship of *CR comment*-metadata in terms of distinguishing our prediction based useful/not-useful annotations. The column 'all' indicates the entire dataset $\mathcal{D}_{sci}$ and P1-10 are ten GitHub SciOSS projects. Though categorical *author association* and *country* metadata could distinguish usefulness for individual projects with small effect sizes, overall, they could not distinguish for the entire dataset, $\mathcal{D}_{sci}$. On the other hand, approximated *gender* metadata shows a very small but consistent association with *CR comment*-usefulness across machine learning, applied mathematics, and programming systems projects, P1-10. In Table 3, we also notice non-categorical metadata *start line* moderately correlates with not-useful, whereas *line, position* with useful *CR comments*. The emoji reactions 👍 show a small correlation in P1 and P7. We also notice that emoji reactions: 👍, 😕, ❤️, and 👀 in P8 can statistically distinguish the usefulness of *CR comments*. Conversely, 👎 only in P10 can statistically distinguish the usefulness.

As we find a small but consistent correlation for gender, country, and author association, developers should be inclusive during code review activities regardless of other participants' origin, identity, and association.

*Feature-data.* Even if the $\mathcal{D}_{sci}$ shows small correlation in Table 1, the individual projects show medium correlation to many features in Table 4. We find negative correlations for *subjectivity, polarity,* and *politeness* features and positive correlations for *CR-sentiment, formality, informativeness,* and *codeword ratio* in P1. The rest of the projects (P2-10) are with similar or lower correlations. While looking at the type of anonymous projects in $\mathcal{D}_{sci}$, we find machine learning related Sci-OSS (P1) is showing stronger feature relations with usefulness than the rest of the applied mathematics (P2-4) and programming systems (P5-10) projects. Since machine learning and deep learning subfields are more widely used by open-source developers, P1 might have a similar feature association to general-purpose *CR comment*($\mathcal{D}_{cc}$-$\mathcal{D}_{rh}$). Additionally, we find in Table 4 that only P7 shows a small correlation with not-useful *CR comment* for features like *refactoring problem keyword density, tone,* and *gratitude.* This intrigues us to compare P7 versus P1-6;8-10 repositories. We find P7 is older than other projects; so, we suspect that the people, community, and culture are different than our other selected projects.

## 4.4 RQ4: Emojis and Usefulness in Scientific *CR comments*

In our derived dataset $\mathcal{D}_{sci}$, we find only 1,269 from 164,708 SciOSS *CR comments* that contain emojis. The $< 1\%$ presence of emojis within *CR comments* in $\mathcal{D}_{sci}$ is much lower than emojis in $\mathcal{D}_{rh}$(2%), $\mathcal{D}_{cc}$(3%), and $\mathcal{D}_{od}$(2.5%). Though the presence of emojis in $\mathcal{D}_{sci}$ is fewer than general, Table 5 shows diverse emojis in *CR comments* from SciOSS. We find emojis in $\mathcal{D}_{sci}$ mostly overlapping with the emojis found in *CR comments* from commercial and open-source

**Table 3: *CR comment*-Usefulness vs *CR comment*-GitHub-Metadata for Various Scientific Software in $\mathcal{D}_{sci}$**

Here, effect-sizes denoted by $*$ are statistically significant (p-value<0.05)

| ↓ **Metadata (Categorical)** | **effect-size**$^{\textbf{p-value}}$ → Cramér's $V^{\text{Chi Square }(\chi^2)}$ | | | | | | | | | | | **Cramér's** $V$ | | | | | | | | | | |
|---|---|---|---|---|---|---|---|---|---|---|---|---|---|---|---|---|---|---|---|---|---|
| $\mathcal{D}_{sci}$-Projects→ | P1 | P10 | P2 | P3 | P4 | P5 | P6 | P7 | P8 | P9 | all | P1 | P10 | P2 | P3 | P4 | | P6 | P7 | P8 | P9 | all |
| author_association | .08* | .02 | .03* | 0* | .03* | 0 | .01* | .05 | .03* | .01* | **.01** | .08 | .02 | .03 | .00 | .03 | | .01 | .05 | .03 | .01 | **.01** |
| country | .08* | .07* | .09 | 0* | .01* | 0 | .06* | 0* | .06* | .03* | **.06** | .08 | .07 | .09 | .00 | .01 | | .06 | | .06 | .03 | **.06** |
| gender | .08* | .03* | .06* | .03* | .03* | 0 | .02* | .13* | .02* | .05* | **.02*** | .08 | .03 | .06 | .03 | .03 | | .02 | .13 | .02 | .05 | **.02** |
| project | 0* | 0* | 0* | 0 | 0* | 0 | 0* | 0 | 0* | 0* | **.05*** | | | | | | | | | | | **.05** |
| side | 0* | .03* | .02* | .05 | .01* | 0 | .04* | 0* | .04* | .02* | **.03*** | .00 | .03 | .02 | .05 | .01 | | .04 | .00 | .04 | .02 | **.03** |
| start_side | 0 | 0* | .02 | .28* | .03* | 0 | .07 | 0 | .06* | .02* | **.05*** | | .00 | .02 | .28 | .03 | | .07 | | .06 | .02 | **.05** |

| ↓ Metadata (Numeric) | **effect-size**$^{\textbf{p-value}}$ → Cohen's $D^{\text{Mann-Whitney U}}$ | | | | | | | | | | | **Pearson** ($\rho$) | | | | | | | | | | |
|---|---|---|---|---|---|---|---|---|---|---|---|---|---|---|---|---|---|---|---|---|---|
| $\mathcal{D}_{sci}$-Projects→ | P1 | P10 | P2 | P3 | P4 | | P6 | P7 | P8 | P9 | all | P1 | P10 | P2 | P3 | P4 | | P6 | P7 | P8 | P9 | all |
| line | .62 | .03 | .03 | .07 | .06 | 0 | .04 | .88 | .03* | .03 | **.03** | .14 | .01 | -.01 | .03 | -.03 | | .02 | .31 | .01 | -.01 | .01 |
| original_line | .56 | .01 | .02 | .02 | .08* | 0 | .01 | .15 | .01 | .02 | **.02** | -.14 | .01 | -.01 | -.01 | -.04 | | .00 | .07 | .00 | .01 | .01 |
| original_position | .1 | .05* | .05* | .11* | .02 | 0 | .03* | .21 | 0* | .08* | **.01*** | .03 | .02 | .02 | -.05 | -.01 | | .01 | -.09 | .00 | .03 | .00 |
| original_start_line | 0 | .12 | .07 | .94 | .1* | 0 | .01 | 0 | .01* | .09 | **.03** | | .05 | .03 | -.37 | -.04 | | -.01 | .27 | .00 | .04 | .01 |
| position | .63 | .04 | .01* | .12 | .04 | 0 | .06* | .48 | .02* | .1* | **.01*** | .15 | .02 | .01 | -.06 | -.02 | | .03 | .18 | .01 | .04 | .01 |
| pr_num | .17 | .01 | .04* | .06 | .02 | 0 | .01 | .04 | 0 | .01 | **.09*** | -.04 | .01 | -.02 | .03 | -.01 | | .00 | -.02 | .00 | .00 | .04 |
| react_+1 | .4 | .02 | .01 | 0 | .03 | 0 | .03 | .17 | .04* | .06* | **.04*** | .10 | -.01 | -.01 | .00 | .01 | | .01 | .07 | .02 | .03 | .02 |
| react_-1 | 0 | .08* | 0 | 0 | 0 | 0 | .01 | 0 | .01 | .01 | **.01*** | | -.03 | | | | | .00 | | -.01 | .00 | .00 |
| react_confused | 0 | .05 | .01 | 0 | .01 | 0 | .02 | 0 | .02* | .01 | **.02*** | | -.02 | .01 | | .00 | | -.01 | | -.01 | .00 | -.01 |
| react_eyes | 0 | .03 | .02 | 0 | .02 | 0 | .01 | 0 | .03* | 0 | **.02*** | | .01 | -.01 | | -.01 | | .00 | | -.01 | | -.01 |
| react_heart | 0 | .01 | .03 | 0 | 0 | 0 | .01 | 0 | .03* | .01 | **.02*** | | .01 | .01 | | | | .01 | | -.01 | .00 | -.01 |
| react_hooray | 0 | .03 | .02 | 0 | .05 | 0 | .01 | 0 | 0 | .02 | **0** | | -.01 | .01 | | -.02 | | .00 | | .00 | .01 | .00 |
| react_laugh | 0 | 0 | .02 | .04 | 0 | 0 | .02 | 0 | .01 | .01 | **.01** | | .00 | .01 | .02 | .00 | | .01 | | .00 | .01 | .00 |
| react_rocket | 0 | .01 | .01 | 0 | .02 | 0 | .02 | 0 | 0 | 0 | **0** | | .01 | .01 | | -.01 | | -.01 | | .00 | | .00 |
| react_total_count | .4 | .03 | 0 | 0 | .02 | 0 | .02 | .17 | .03* | .07* | **.03*** | .10 | -.01 | .00 | .00 | .01 | | .01 | .07 | .01 | .03 | .01 |
| start_line | 0 | .11 | .05 | .99 | .09 | 0 | .1 | 0 | .03 | .05 | **.03** | | .05 | -.02 | -.39 | -.04 | | .04 | | .01 | .02 | .01 |

general software. Interestingly, we did not find any customized GitHub emojis, :shipit: or :trollface:, from 164,708 Sci-OSS *CR comments*. However, we see a few new emojis with sparse presence, such as - 😓(4) 🐎(3) 😸(2) 😭(2) 💀(1) ☺️(1) ✌️(1) 👕(1) 🤢(1) 🐙(1).

Table 6 shows the metadata and features' correlation with usefulness of *CR comments* on the entire dataset $\mathcal{D}_{sci}$ and emoji-containing *CR comments* in $\mathcal{D}_{sci}$, denoted by $\mathcal{D}_{sci}^{\in}$🙂. The notable differences in correlation we see for metadata are *country, start side,* and *start line*, and for features are *polarity, subjectivity, stop word ratio,* and *emoji sentiment*. Among these metadata and features, we see that only the *start line* has the most contrasting difference. Therefore, emoji-containing *CR comments* in $\mathcal{D}_{sci}$ are slightly different. Next, we look into the semantic similarity or difference between $\mathcal{D}_{sci}$ and $\mathcal{D}_{sci}^{\in}$🙂.

Table 8 shows stratified 10-fold cross-validation *CR comment*-usefulness prediction performance with emoji-containing *CR comments* in $\mathcal{D}_{sci}$, $\mathcal{D}_{sci}^{\in}$🙂. Though emoji semantics helped in predicting the usefulness of *CR comments* better in general software [2], Table 8 shows that emoji-semantics-aware models could not predict usefulness better than emoji-semantics-non-aware models. With our additional statistical measure, we find that all approaches reject the Wilcoxon signed-rank null hypothesis. That is, the paired samples of the original and predicted usefulnesses are significantly different. Also, the Cohen's D effect sizes resemble the results from traditional classification metrics.

Ahmed & Eisty also found that emoji semantics performed poorly when $\mathcal{D}_{rh}^{\in}$🙂, $\mathcal{D}_{cc}^{\in}$🙂, and $\mathcal{D}_{od}^{\in}$🙂 were unified for usefulness prediction with semantics and 10-fold cross-validation. Since $\mathcal{D}_{sci}$ comprises 10 prominent SciOSS, we also infer that emojis are used and perceived differently within these 10 SciOSS projects. To verify our inference, we look up the emojis in $\mathcal{D}_{sci}^{\in}$🙂 and their associated automated usefulness labels (Section 3.1.2). We present 10 SciOSS projects of $\mathcal{D}_{sci}$, the emojis in the *CR comments*, and their associated usefulness labels in Table 7. Within $\mathcal{D}_{sci}$-projects P1-10, we see that P1 and P5 do not have emoji-containing *CR comments*. We also notice that most emojis belong to both useful and not-useful classes. For cross projects, we also find different distributions of emojis. Thus, the models learned different representations of the same emojis when emoji embedding was aggregated with text embedding.

Though the emojis in SciOSS have a sparser presence and show variation in usage within and across our selected SciOSS repositories, emoji reactions on code-reviewers' *CR comments* also correlate with usefulness. So, the code-author should meaningfully react with emojis, and the code-reviewer should conscientiously perceive emoji reactions to their feedback.

## 4.5 Practical Implications

Our analysis offer the following implication for code-reviewers and code-authors of Sci-OSS.

- Our findings confirm prior findings [3] on general software. Therefore, code-reviewers in Sci-OSS should compose positive, respectful, and objective *CR comments* as well. They also need to specify source-code lines while providing feedback.
- Since developers pay more attention to emojis than text and they carry instructions and emotions, emojis should be

**Table 4: Statistical Properties of Features in $\mathcal{D}_{sci}$ *CR comments***

Here, effect-sizes denoted by $*$ are statistically significant (p-value<0.05)

| | effect-size$^{\text{p-value}}$ | | | | | | | | | | Pearson Correlation($\delta$) | | | | | | | | | |
|---|---|---|---|---|---|---|---|---|---|---|---|---|---|---|---|---|---|---|---|---|
| $\mathcal{D}_{sci}$-*Projects*→ | P1 | P10 | P2 | P3 | P4 | P6 | P7 | P8 | P9 | all | P1 | P10 | P2 | P3 | P4 | P6 | P7 | P8 | P9 | all |
| tone | .24 | .04* | .04 | .02 | .03 | .13* | .33 | .08* | .03 | .03* | -.06 | .02 | -.02 | -.01 | .01 | -.05 | -.13 | .03 | -.01 | .01 |
| distress | .07 | .04* | .03 | .05 | .01 | .07 | .13 | .01* | 0* | .01* | -.02 | .02 | -.01 | .02 | .00 | -.03 | -.05 | .00 | .00 | .00 |
| empathy | .07 | .04* | .03 | .05 | .01 | .07 | .13 | .01* | 0* | .01* | -.02 | .02 | -.01 | .02 | .00 | -.03 | -.05 | .00 | .00 | .00 |
| is_toxic | .14 | .02 | .04 | .03 | .03 | .02 | 0 | .01 | .02 | .01 | .03 | -.01 | -.02 | .01 | .01 | -.01 | | .00 | .01 | .00 |
| gratitude | .1 | .02 | .06* | .1* | .02 | .03 | .18 | .02* | .03 | 0 | .02 | -.01 | -.03 | -.05 | .01 | -.01 | -.07 | .01 | -.01 | .00 |
| cr_senti | 1.29* | .26* | .22* | .3* | .25* | .36* | .03 | .31* | .15* | .29* | .30 | .11 | .10 | .14 | .11 | .15 | -.01 | .13 | .06 | .12 |
| kw_refact_xerox | .36 | .19* | .15* | .19* | .12* | .15* | .19 | .15* | .21* | .15* | .09 | .08 | .06 | .09 | .05 | .06 | .08 | .06 | .09 | .06 |
| num_propernouns | 0 | .04 | .01 | .11* | .02 | .02 | 0 | .01* | .01 | .01* | | -.02 | .01 | -.05 | -.01 | .01 | | .00 | .00 | .00 |
| num_interjections | 0 | .03 | .03 | .11* | .01 | .01 | 0 | 0 | 0 | 0 | | -.01 | .01 | -.05 | .00 | .00 | | .00 | .00 | .00 |
| polarity | 1.27* | .27* | .3* | .27* | .22* | .33* | .4 | .25* | .36* | .27* | -.30 | -.11 | -.13 | -.13 | -.10 | -.14 | -.16 | -.10 | -.15 | -.11 |
| kw_msoft_u | .23 | .26* | .21* | .18* | .15* | .18* | .34 | .21* | .24* | .21* | .06 | .11 | .09 | .08 | .07 | .08 | .14 | .08 | .10 | .09 |
| num_exclamation | .2 | .05* | .05* | .13* | .02 | .03 | .17 | .02* | 0 | .01* | .05 | .02 | -.02 | -.06 | .01 | .01 | -.07 | .01 | .00 | .01 |
| is_confirmatory | .33 | .18* | .21* | .19* | .12* | .17* | .17 | .24* | .24* | .22* | .08 | .07 | .09 | .09 | .05 | .07 | .07 | .10 | .10 | .09 |
| question_ratio | .24 | .26* | .15* | .14* | .2* | .21* | .1 | .21* | .06* | .19* | -.06 | -.11 | -.06 | -.06 | -.09 | -.09 | .04 | -.08 | -.02 | -.08 |
| kw_satd_potdar | .35 | .04 | .05* | 0 | .04 | .04* | .19 | .01 | .07* | .02* | .09 | .02 | .02 | .00 | -.02 | .02 | .08 | .00 | .03 | .01 |
| implicature | .19 | 0 | .02 | .06 | .11* | .04 | .17 | .13* | .05 | .09* | -.05 | .00 | .01 | -.03 | -.05 | -.01 | -.07 | -.05 | -.02 | -.04 |
| kw_msoft_nu | .2 | .18* | .16* | .27* | .1* | .29* | .09 | .21* | .17* | .21* | .05 | -.08 | -.07 | -.13 | -.05 | -.12 | .04 | -.08 | -.07 | -.09 |
| code_word_ratio | .55 | .16 | .13* | .16* | .19* | .27* | .05 | .19* | .16* | .2* | .13 | .07 | .06 | .08 | .08 | .11 | .02 | .08 | .07 | .08 |
| politeness | 1.07 | .15* | .23* | .18* | .16* | .13* | .14 | .12* | .26* | .14* | -.25 | -.06 | -.10 | -.09 | -.07 | -.05 | -.06 | -.05 | -.11 | -.06 |
| num_Qmark | .87* | .26* | .11* | .15* | .17* | .22* | .15 | .11* | .08* | .12* | -.21 | -.11 | -.05 | -.07 | -.07 | -.09 | .06 | -.04 | -.03 | -.05 |
| rd_text | .8* | .2* | .21* | .29* | .2* | .31* | .28 | .19* | .14* | .21* | -.19 | -.08 | -.09 | -.13 | -.09 | -.13 | -.11 | -.07 | -.06 | -.08 |
| kw_refact_problem | .34 | .04 | .13* | .01 | .01 | .07 | .45* | 0 | .05* | .01 | .08 | .02 | -.05 | .00 | .00 | -.03 | -.18 | .00 | .02 | -.01 |
| avg_chars | .83* | .19* | .19* | .27* | .19* | .27* | .3 | .2* | .13* | .2* | .20 | .08 | .08 | .13 | .08 | .11 | .12 | .08 | .06 | .08 |
| formality | .88 | .09* | .13* | .16* | .09* | .15* | .19 | .07* | .17* | .1* | .21 | .04 | .06 | .07 | .04 | .06 | .08 | .03 | .07 | .04 |
| subjectivity | 1.5* | .34* | .31* | .25* | .34* | .35* | .47* | .38* | .3* | .36* | -.34 | -.14 | -.13 | -.12 | -.15 | -.14 | -.19 | -.15 | -.13 | -.15 |
| has_out_snippet | .06 | .04 | 0 | .1* | .02 | .22* | .09 | .02* | .04 | .04* | -.01 | -.02 | .00 | .05 | -.01 | .09 | -.04 | .01 | .02 | .02 |
| kw_secdev | .05 | .16* | .03 | .03 | .03* | .05* | .1 | .16* | .03 | .12* | .01 | -.07 | -.01 | -.02 | -.01 | -.02 | -.04 | -.06 | -.01 | -.05 |
| avg_punct | .37 | .07 | .08* | .24* | .11* | .16* | .17 | .1* | .08 | .1* | .09 | .03 | .03 | .11 | .05 | .07 | .07 | .04 | .04 | .04 |
| num_sent | .16 | .28* | .12* | .16* | .23* | .18* | .05 | .23* | .13* | .21* | -.04 | -.11 | -.05 | -.07 | -.10 | -.07 | -.02 | -.09 | -.06 | -.09 |
| stop_word_ratio | .51 | .32* | .24* | .16* | .32* | .42* | .19 | .4* | .23* | .37* | -.12 | -.13 | -.10 | -.07 | -.14 | -.17 | -.08 | -.16 | -.09 | -.15 |
| num_tentative | .18 | .18* | .06* | .07 | .1* | .06* | .01 | .22* | .11* | .17* | .04 | -.08 | -.03 | -.03 | -.05 | -.02 | -.01 | -.09 | -.05 | -.07 |
| informativeness | .7 | .04 | .14* | .1* | .04 | .13* | .05 | .01 | .17* | .06* | .17 | .02 | .06 | .04 | .02 | .05 | .02 | .01 | .07 | .02 |
| programming_words | .22 | .03* | .03 | .08 | .05 | .1* | .09 | 0* | .08* | .02 | .05 | -.01 | .01 | .04 | .02 | .04 | .04 | .00 | .03 | .01 |
| num_adverb | .4 | .28* | .12* | .09* | .18* | .25* | .09 | .32* | .12* | .27* | -.10 | -.11 | -.05 | -.04 | -.08 | -.10 | -.04 | -.13 | -.05 | -.11 |
| num_adj | .08 | .21* | .06* | .15* | .08* | .04* | .02 | .16* | .11* | .13* | .02 | -.09 | -.03 | -.07 | -.03 | -.02 | -.01 | -.06 | -.04 | -.05 |
| num_nouns | .39 | .15* | .02 | .05 | .02* | .06* | .06 | .09* | .02* | .06* | .09 | -.06 | .01 | .02 | -.01 | .02 | .03 | -.04 | -.01 | -.02 |
| num_determinants | .19 | .29* | .12* | .12* | .23* | .2* | .11 | .31* | .12* | .26* | .05 | -.12 | -.05 | -.06 | -.10 | -.08 | -.05 | -.12 | -.05 | -.11 |
| num_chars | .21 | .21* | .01* | .02 | .04* | .01* | .03 | .16* | .05* | .11* | .05 | -.09 | -.01 | -.01 | -.02 | .01 | .01 | -.06 | -.02 | -.05 |
| num_words | .05 | .27* | .06* | .09* | .14* | .11* | 0 | .25* | .1* | .2* | .01 | -.11 | -.03 | -.04 | -.06 | -.04 | .00 | -.10 | -.04 | -.08 |
| avg_stopwords | .07 | .28* | .15* | .08 | .19* | .28* | .14 | .35* | .14* | .29* | -.02 | -.12 | -.06 | -.04 | -.08 | -.12 | -.06 | -.14 | -.06 | -.12 |
| kw_refact_solution | .33 | .26* | .05* | .03 | .16* | .16* | .02 | .22* | .06* | .19* | .08 | -.11 | -.02 | -.01 | -.07 | -.07 | -.01 | -.09 | -.03 | -.08 |
| num_verb | .27 | .2* | .01 | .01 | .13* | .06* | .03 | .21* | .04 | .16* | -.07 | -.08 | .00 | -.01 | -.06 | -.03 | -.01 | -.09 | -.02 | -.07 |
| avg_words | .11 | .16* | .03* | .02 | .01* | .02* | .02 | .16* | .02* | .11* | .03 | -.07 | -.01 | -.01 | .00 | -.01 | -.01 | -.06 | -.01 | -.04 |

**Table 5: Emojis in *CR Comments***

Here, numbers inside the parentheses following the emojis denote the appearance of the emojis

| **Dataset** | **Emojis in the Dataset** |
|---|---|
| $\mathcal{D}_{rh}$ [27] (1,481) | 👍(18), 😄(9), 😞(3), 😉(2), 😳(1), 👎(1), 😲(1), 🙍(1), 😛(1), 🙅(1), :shipit: (1), :trollface:(1) |
| $\mathcal{D}_{cc}$ [24] (3,794) | 😄(54), 😉(6), 😞(3), 🤓(2), 😕(1), 😛(1), 🕶(1). |
| $\mathcal{D}_{od}$ [31] (2,654) | 😄(38), 😞(8), 😛(3), 👍(2), 😉(3) |
| $\mathcal{D}_{sci}$ [Ours] (164,708) | 😉(290) 👍(257) 😞(210) 😅(85) 🙂(79) 😕(66) 🕶(62) 🤔(47) 😛(43) 😄(28) 🤷(23) 😃(19) 🤦(17) 😆(15) 😁(12) 😦(9) 🙃(8) 😜(8) 🙄(6) 😢(5) 😲(5) 😀(4) 😓(4) 😔(3) 🐎(3) 👀(3) 😳(3) 😕(3) 😝(2) 😸(2) 😭(2) 💀(1) 😂(1) 🤓(1) ☺️(1) 😨(1) 🤞(1) 👕(1) 🤢(1) 🐙(1) |

**Table 6: Metadata and Features' Correlation in $\mathcal{D}_{sci}^{\in}$🙂**

| **Metadata** | $\mathcal{D}_{sci}$ | $\mathcal{D}_{sci}^{\in}$🙂 | **Features** | $\mathcal{D}_{sci}$ | $\mathcal{D}_{sci}^{\in}$🙂 | | $\mathcal{D}_{sci}$ | $\mathcal{D}_{sci}^{\in}$🙂 |
|---|---|---|---|---|---|---|---|---|
| author_association | **.01** | **.05** | tone | .01 | .05 | programming_words | .01 | -.02 |
| country | **.06** | **.19** | distress | .00 | .00 | num_adverb | -.11 | -.13 |
| gender | **.02** | **.09** | empathy | .00 | .00 | num_adj | -.05 | -.03 |
| project | **.05** | **.04** | is_toxic | .00 | .00 | num_nouns | -.02 | -.03 |
| side | **.03** | **.03** | gratitude | .00 | .00 | num_determinants | -.11 | -.11 |
| start_side | **.05** | **.13** | cr_senti | .12 | .08 | num_chars | -.05 | -.05 |
| line | .01 | **-.01** | kw_refact_xerox | .06 | .03 | num_words | -.08 | -.08 |
| original_line | **.01** | **.01** | num_propernouns | .00 | -.02 | emo_senti | -.01 | .034 |
| original_position | .00 | **.02** | num_interjections | .00 | .03 | avg_stopwords | -.12 | -.16 |
| original_start_line | .01 | **.06** | polarity | -.11 | -.17 | kw_refact_solution | -.08 | -.08 |
| position | .01 | **.00** | kw_msoft_u | .09 | .05 | num_verb | -.07 | -.08 |
| pr_num | .04 | **.00** | num_exclamation | .01 | .01 | avg_words | -.04 | -.04 |
| react_+1 | .02 | **.03** | is_confirmatory | .09 | .06 | subjectivity | -.15 | -.20 |
| react_-1 | .00 | | question_ratio | -.08 | -.01 | has_out_snippet | .02 | -.03 |
| react_confused | -.01 | | kw_satd_potdar | .01 | -.06 | kw_secdev | -.05 | -.08 |
| react_eyes | -.01 | **-.02** | implicature | -.04 | -.02 | avg_punct | .04 | .04 |
| react_heart | -.01 | **-.05** | kw_msoft_nu | -.09 | .02 | num_sent | -.09 | -.09 |
| react_hooray | .00 | **-.02** | code_word_ratio | .08 | .02 | stop_word_ratio | -.15 | -.19 |
| react_laugh | .00 | **.05** | politeness | -.06 | -.09 | num_tentative | -.07 | -.06 |
| react_rocket | .00 | **.04** | num_Qmark | -.05 | .00 | informativeness | .02 | .05 |
| react_total_count | .01 | **.04** | rd_text | -.08 | .05 | avg_chars | .08 | .03 |
| start_line | .01 | **-.17** | kw_refact_problem | -.01 | .03 | formality | .04 | .09 |

**Table 7: Emoji Presence in *CR comments* within $\mathcal{D}_{sci}$**

Here, numbers inside the parentheses following the emojis denote the appearance of the emojis

| $\mathcal{D}_{sci}$ | **Emojis in Useful *CR comments*** | **Emojis in Not-useful *CR comments*** |
|---|---|---|
| P2 | 😉(39) 👍(30) 😞(13) 😛(13) 😜(6) 🐎(3) 😅(3) 😄(2) 😃(2) 😕(1) 🐙(1) 👀(1) 😁(1) 🤔(1) 🙂(1) | 😉(26) 😞(5) 👍(3) 😄(2) 😕(2) 😛(2) 😝(1) 😃(1) 😁(1) |
| P3 | 😞(1) | 😞(1) |
| P4 | 🤷(6) 👍(4) 🤔(2) 😆(2) 😉(2) 🤦(1) 👀(1) 😞(1) 😁(1) 😃(1) 😅(1) 😕(1) 🕶(1) 🙄(1) | 👍(3) 😅(2) 🕶(2) 😉(1) 😁(1) 😞(1) 😕(1) 🤷(1) |
| P6 | 😉(39) 🙂(33) 👍(22) 😕(17) 😞(9) 😛(5) 🕶(5) 🤔(3) 🙄(2) 🤦(2) 😁(1) 🤞(1) 🤷(1) 🤓(1) 😅(1) 🙁(1) 😆(1) | 😉(13) 🙂(13) 😞(6) 😕(6) 🤷(2) 😄(1) 🙁(1) 😁(1) 😛(1) 🕶(1) |
| P8 | 😞(113) 😉(107) 👍(73) 😅(52) 🕶(34) 🤔(28) 🙂(26) 😕(22) 😛(14) 😄(10) 🤷(7) 🤦(7) 🙃(6) 😃(5) 😆(4) 😔(3) 😭(2) 😁(2) 😢(2) 👀(1) 😂(1) 😀(1) 😸(1) 😓(1) 😨(1) | 😞(37) 😉(36) 😅(26) 👍(16) 😕(13) 🕶(11) 🤔(10) 😄(9) 😛(7) 🙂(5) 🤦(3) 😓(3) 🤷(3) 😃(2) 🙃(2) 😳(1) 🙄(1) 😢(1) 😀(1) 😁(1) ☺️(1) 😆(1) |
| P9 | 👍(58) 😞(6) 😆(5) 😉(5) 😃(4) 🤦(3) 🤷(3) 😁(2) 🕶(2) 💀(1) 👕(1) 🙄(1) 😸(1) 🤢(1) 😦(1) 🙂(1) 😀(1) | 👍(17) 😞(3) 😃(3) 🕶(3) 😆(1) 😀(1) 😁(1) 🤔(1) |
| P10 | 👍(22) 😉(17) 😞(8) 😲(5) 😦(4) 🕶(3) 😕(2) 😄(2) 🤔(2) 😜(2) 😛(1) 😃(1) 😳(1) 🙁(1) 😢(1) 🤦(1) | 👍(9) 😞(6) 😉(5) 😦(4) 😄(2) 😕(1) 😢(1) 😳(1) 😆(1) 😝(1) |

**Table 8: Emoji-Semantics in Identifying Usefulness of *CR comments* from Scientific Software**

10-fold Stratified Cross-validation Performance of models with emojis and without emojis.

| | **P** | **R** | $F_1$ | **M** | **A** | $D^p$ |
|---|---|---|---|---|---|---|
| $\mathcal{D}_{sci}^{\in}$🙂 (text) | .76 | .99 | .86 | .22 | .76 | $.64^*$ |
| $\mathcal{D}_{sci}^{\in}$🙂 (emoji) | .74 | .98 | .84 | -.02 | .73 | $.73^*$ |
| $\mathcal{D}_{sci}^{\in}$🙂 (text+emoji) | .75 | .99 | .85 | .13 | .75 | $.69^*$ |

Here, **P**: Precision, R: Recall, **A**: Accuracy, $F_1$: $F_1$-scores, **M**: Matthews correlation coefficient (MCC), $D^p$: Cohen's D effect size$^{\text{Wilcoxon signed-rank p-value}}$, and $*$ : p-value $< 0.05$ carefully used similar to words [2]. For emoji-containing *CR comments* in scientific software, we find more association with usefulness, so it reinforces significance of emoji usage. Thus, code-authors should meaningfully react with emojis. Also, code-reviewers should perceive emoji reactions on their feedback conscientiously.

# 5 THREATS TO VALIDITY

Our experiment setup leverages the existing theories, datasets, experiments, and approaches for identifying and analyzing useful *CR comments*. However, we discuss the potential threats to the validity of our work.

## 5.1 External Validity

The generalizability of our findings is a threat to external validity. Our experiment data are from a small set of Sci-OSSs from applied mathematics, programming systems, and machine learning domains. We expect our findings to apply to Sci-OSS broadly; however, there remains a threat to external validity for projects from other domains.

## 5.2 Internal Validity

First, model computed metrics in our work are trained from different sources; thus, they may carry implicit biases that we are not aware of since the data was collected by other researchers. Second, our approximated gender values are unknown for meaningless or encrypted names, and normalized location values are missing for GitHub profiles sharing improper information. Our findings regarding gender or location could have been different if these data points had been available directly from the GitHub profiles. Third, Developers may communicate through alternative channels (e.g., email, Discord, Slack), which may not be captured in *CR comments*. However, our work evaluates the usefulness of *CR comments* based exclusively on their textual content, rather than the complete communication process. Therefore, this does not constitute a significant threat to internal validity of our findings.

## 5.3 Conclusion Validity

To conclude our findings accurately, we carefully designed our experiments. We compared feature values of our prediction-based useful *CR comments* with existing datasets. Since the state-of-the-art prediction model we used for annotation is not from handcrafted feature-based model, analyzing the features for comparison does not impose any threats. In our statistical analysis, we have considered suitable statistical tests for numeric and categorical metadata/features so that our findings do not violate the underlying normality and independence assumptions.

# 6 CONCLUSIONS

Researchers have analyzed, characterized, and identified the usefulness of *CR comments* in non-scientific closed and open-source software projects. In this paper, we investigate the *CR comments* in scientific open-source software to describe the distinctive nature of scientific *CR comments*. We leveraged existing usefulness-annotated *CR comment* datasets, the best-performing usefulness prediction model, and feature analysis techniques. Our statistical and XAI feature analyses confirm findings from prior research on general-purpose software. With our additional derived metadata, we analyzed their association with the usefulness of code review feedback. These findings will benefit developers, leaders, scientists, and researchers in both general and Sci-OSS communities. In the future, we plan to conduct an empirical study on the usefulness *CR comments* focusing scientific open source software code authors and code reviewers.

# 7 DATA AVAILABILITY

We have shared our code and dataset, including feature values computed in our experiments at
https://github.com/sharif509/SciOSS-CRCs-artifact